\input{aipcheck}

\documentclass[
    ,final            
  ]
  {aipproc}

\layoutstyle{6x9}

\begin{document}
\def\be{\begin{equation}}
\def\ee{\end{equation}}
\def\bea{\begin{eqnarray}}
\def\eea{\end{eqnarray}}

\title{The geometry of thermodynamics}


\author{Hernando Quevedo}{
  address={Instituto de Ciencias Nucleares, 
Universidad Nacional Aut\'onoma de M\'exico  \\
 A.P. 70-543, 04510 M\'exico D.F., MEXICO}
}

\author{Alejandro V\'azquez}{
  address={Facultad de Ciencias, 
Universidad Aut\'onoma del Estado de Morelos \\
Av. Universidad 1001,   
 Cuernavaca 62210, Morelos, MEXICO}
}

\begin{abstract}
 We present a review of the main aspects of geometrothermodynamics, 
an approach which allows us to associate a specific Riemannian structure 
to any classical thermodynamic system. In the space of equilibrium states,
we consider a Legendre invariant metric, which is given in terms of the 
fundamental equation of the corresponding thermodynamic system, and analyze 
its geometric properties in the case of the van der Waals gas, and
black holes. We conclude that the geometry of this particular metric 
reproduces the thermodynamic
behavior of the van der Waals gas, and the Reissner-Nordstr\"om black hole,
but it is not adequate for the thermodynamic description of Kerr black holes. 
\end{abstract}

\maketitle


\section{Introduction}
\label{sec:int}

Differential geometry is a very important tool of modern science, 
specially of mathematical physics and its applications in physics,
 chemistry and engineering. 
In the case of thermodynamics, Gibbs \cite{gibbs} and 
Caratheodory \cite{car}, followed by Hermann \cite{her} and later by Mrugala
\cite{mru1,mru2}, proposed a differential geometric approach based 
upon the contact structure of the thermodynamic phase space ${\cal T}$.
This space is $(2n+1)-$dimensional and is coordinatized by $n$ extensive 
variables $E^a$ and $n$ intensive variables $I^a$, together with 
the thermodynamic potential $\Phi$. The first law of thermodynamics
is incorporated into this approach in a very natural way through differential 
forms. A particular $n-$dimensional subspace of ${\cal T}$ is the space of
thermodynamic equilibrium states ${\cal E}$. On ${\cal E}$, the laws of 
thermodynamics are valid and thermodynamics systems are specified by means
of a fundamental equation. 

In an attempt to describe thermodynamic systems in terms of geometric objects,
Weinhold \cite{wei1} introduced {\it ad hoc} on the space of equilibrium states
 a metric whose components are given as the Hessian of the internal 
thermodynamic energy. 
This metric 
turns out to be positive as a consequence of the second law of 
thermodynamics. This approach has been intensively used to study, from
a geometrical point of view, the properties of the space generated 
by Weinhold's metric \cite{fel1,gil1}, the thermodynamic length
\cite{sal80,sal84,sal85}, the chemical and physical
properties of various two-dimensional thermodynamic systems 
\cite{nul85,san04,san05a,san05b,san05c}, and the associated 
Riemannian structure \cite{rup79,tormon93,herlac98}. 

In an attempt to understand the concept of thermodynamic length, Ruppeiner
\cite{rup79}
introduced a metric which is given as the Hessian of the entropy 
and is conformally equivalent to Weinhold's metric,
with the inverse of the temperature as the conformal factor. 
The physical meaning of Ruppeiner's geometry lays in the fluctuation 
theory of equilibrium thermodynamics. It turns out that the second
moments of fluctuation are related to the components of the inverse of
Ruppeiner's metric. The corresponding geometry has been investigated for
several thermodynamic systems such as the ideal (classic and quantum) gas,
 one-dimensional Ising model, multicomponent ideal gas, van der Waals gas, etc.
It was shown that Ruppeiner's metric contains important information 
about the phase transition structure of thermodynamic systems, indicating
the location of critical points and phase transitions on those particular
surfaces where the scalar curvature diverges. In the case of systems with 
no statistical mechanical interactions (e.g. an ideal gas), the scalar
curvature vanishes and consequently the geometry of the associated 
two-dimensional space is flat. For this reason the thermodynamic length
is considered  as a measure of statistical mechanical interactions
within a thermodynamic system. These results have been reviewed in
\cite{rup95} and more recent results are included in 
\cite{jjk03,jan04}. Due to the conformal equivalence, Weinhold's metric 
contains similar information about the structure of phase transitions
as has been shown recently in \cite{san05c}. Ruppeiner's metric has found
applications also in the context of thermodynamics of black holes
 \cite{aman03,aman06a,aman06b,shen05}. 

An alternative approach has been used in classical statistical mechanics
to analyze the geometry of thermodynamic systems. The starting 
point is the probability density distribution which includes the 
partition function of the corresponding system. It can be shown \cite{brohug1,
brohug2,brohug3} that from the information contained in the partition
function a metric structure can be derived, the so-called Fisher-Rao metric
which describes the properties of the statistical system. Although their 
conceptual origin is quite different, it can be shown \cite{jjk03}  
that Weinhold and Ruppeiner metrics are related to the Fisher-Rao metric by means 
of Legendre transformations of the corresponding thermodynamic variables. 
One common disadvantage of all these metrics is that they are not Legendre 
invariant, leading to the unphysical result that their properties depend
on the thermodynamic potential used in their construction. 

Recently, the formalism of geometrothermodynamics (GTD) was developed in \cite{quezar03,quev07}
in order to incorporate the concept of Legendre invariance into the geometric
description of thermodynamics. It unifies in a consistent manner the contact 
structure of ${\cal T}$ with the metric structures on ${\cal E}$. One of the 
main results of GTD is that it allows to introduce Legendre invariant metrics
on ${\cal T}$ which are then consistently projected on ${\cal E}$, generating
in this way metrics which can be used to describe the properties of 
thermodynamic systems in terms of geometric properties. 

In this work, we present a simple Legendre invariant metric which is then applied
to different thermodynamic systems in order to analyze their geometric properties.
We first present a short review of the main constituents of
GTD, and consider a specific Legendre invariant metric. Then, 
we study the geometric properties of the metric corresponding to 
the van der Waals gas and to the ideal gas, 
as a special case of the first one. 
It follows that the geometry 
reproduces the thermodynamic behavior in both cases.
Furthermore, we consider black holes as
thermodynamic systems and analyze their geometric properties. 
The last section of this work is devoted
to discussions of our results and suggestions for further research.

\section{Geometrothermodynamics}
\label{sec:gtd}

Let the thermodynamic phase space ${\cal T}$ be
coordinatized by the thermodynamic potential $\Phi$, extensive variables $E^a$, 
and intensive variables $I^a$ $(a=1,...,n)$. 
Let the fundamental Gibbs 1-form be defined on ${\cal T}$ as 
$\Theta = d\Phi - \delta_{ab} I^a d E^b$, with $ 
\delta_{ab}={\rm diag} (1,1,...,1)$, satisfying the condition 
$\Theta \wedge (d\Theta)^n \neq 0$. Consider also on ${\cal T}$
the Riemannian metric
\be
G = (d\Phi - \delta_{ab} I^a d E^b)^2 
+ (\delta_{ab} E^a I^b)\ (\delta_{cd} d E^c d I^d)   \ ,
\label{genGII}
\ee
which is non-degenerate and invariant with respect to Legendre transformations of the form \cite{arnold}
\be
\{\Phi, E^a,I^a\}\longrightarrow \{\tilde \Phi, \tilde E ^a, \tilde I ^ a\}
\ee
\be
 \Phi = \tilde \Phi - \delta_{kl} \tilde E ^k \tilde I ^l \ ,\quad
 E^i = - \tilde I ^ {i}, \ \  
E^j = \tilde E ^j,\quad   
 I^{i} = \tilde E ^ i , \ \
 I^j = \tilde I ^j \ ,
 \label{leg}
\ee
where $i\cup j$ is any disjoint decomposition of the set of indices $\{1,...,n\}$,
and $k,l= 1,...,i$. In particular, for $i=\{1,...,n\}$ and $i=\emptyset$ we obtain
the total Legendre transformation and the identity, respectively. We say that the 
set $({\cal T}, \Theta, G)$ defines a Legendre invariant manifold with a contact Riemannian 
structure. 

The space of thermodynamic equilibrium states is an $n-$dimensional Riemannian submanifold
${\cal E} \subset {\cal T}$ induced by a smooth mapping 
$ \varphi : \   {\mathcal E} \  \longrightarrow {\mathcal T}$, i.e.
$ \varphi :  (E^a) \longmapsto (\Phi, E^a, I^a)$ 
with $\Phi=\Phi(E^a)$  such that 
\be
\varphi^*(\Theta) = 0\ ,\qquad \varphi^*(G)=g=\Phi \frac{\partial^2\Phi}
{\partial E^a \partial E^b} dE^a dE^b \ , 
\label{metg}
\ee
where $\varphi^*$ is the pullback of $\varphi$ and $g$ is the Riemannian 
metric induced on ${\cal E}$. The first of these equations implies that on 
${\cal E}$ the following relationships hold
\be
d\Phi = \delta_{ab} I^a d E^b \ , \qquad \frac{\partial\Phi}{\partial E^a} = 
\delta_{ab} I^b \ ,
\ee
which correspond to the first law of thermodynamics and the 
conditions for thermodynamic equilibrium, respectively \cite{callen}.
The metric $g$ on ${\cal E}$ is Legendre invariant because it is induced by 
a smooth mapping from the Legendre invariant metric $G$ of ${\cal T}$. 
The explicit components of $g$ can be computed from the fundamental equation 
$\Phi=\Phi(E^a)$, which is given as part of the smooth mapping $\varphi$. 
Since the fundamental equation characterizes completely a thermodynamic system,
the metric $g$ is also a characteristic which is different for each 
thermodynamic system. It is in this sense that we propose to analyze 
the relationship between the thermodynamic properties of a system, 
specified through the fundamental equation $\Phi
=\Phi(E^a)$, and the geometric properties of the corresponding metric $g$.

In general, the thermodynamic potential is a homogeneous function of its arguments, i.e.,
$\Phi(\lambda E^a) = \lambda^\beta \Phi(E^a)$ for constant parameters $\lambda$ and
$\beta$. Using the first law of thermodynamics, it can easily be shown that this homogeneity
is equivalent to  the relationships
\be
\beta \Phi(E^a) = \delta_{ab}I^b E^a \ , \qquad
(1-\beta)\delta_{ab}I^a d E^b +\delta_{ab} E^a d I^b = 0 \ ,
\ee
which are known as Euler's identity and Gibbs-Duhem relation, respectively.
Moreover, the second law of thermodynamics corresponds to 
the convexity condition on the thermodynamic potential 
$ \partial^2 \Phi/\partial E ^a \partial E ^b \geq 0$ \cite{callen,burke}.

It must be noticed that the metric $g$ given in Eq.(\ref{metg}) is not the only one
which is Legendre invariant. In fact, there exists an infinite number of Legendre 
invariant metrics on ${\cal E}$. In this work, however, we limit ourselves to the 
analysis of this specific metric due to its simplicity and the fact that it is the
simplest Legendre invariant generalization of Weinhold's and Ruppeiner's metrics 
\cite{quev07}. 

\section{The van der Waals gas}
\label{sec:vdw}

The van der Waals gas is a 2-dimensional thermodynamic system for which the 
extensive variables can be chosen as the entropy $S$ and the volume $V$. The
corresponding dual intensive variables are the temperature $T$ and pressure $P$. 
Accordingly, the first law of thermodynamics reads $d\Phi = T d S - P d V$, where
the thermodynamic potential $\Phi$ can be identified with the internal energy of 
the gas. The thermodynamic properties of the van der Waals gas can be completely 
derived from the fundamental equation
\be
\Phi(S,V)= \left(\frac   {e^{S/k}}  {V-b}\right)^{2/3}   -\frac{a}{V} \ ,
\label{fevdw}
\ee
where $k$ is a constant, usually related to the Boltzmann constant,  and $a$ and $b$
are constants which are responsible for the thermodynamic interaction between the 
particles of the gas.

On the space of equilibrium states ${\cal E}$, the Riemannian structure is
determined by the metric (\ref{metg}) which in this particular case can be written 
as  
\be
g_{_{vdW}}= \frac{2}{9k^2} \Phi \left(\Phi+\frac{a}{V}\right)\left[ 2 dS^2 - \frac{4k }{V-b}dS dV
+\frac{5k^2 }{(V-b)^2}dV^2\right] -\frac{2a\Phi}{V^3} dV^2 \ .
\ee
This metric defines a 2-dimensional differential manifold whose geometric properties can be 
derived from the analysis of the corresponding connection and curvature. The connection determines
the form of the geodesics in this manifold, and deserves a separate and detailed study which will be
reported elsewhere. The scalar curvature can be computed in a straightforward manner and we obtain
\be
R_{_{vdW}}=\frac{a{\cal P}(\Phi,V,a,b)}{\Phi^3(PV^3-aV+2ab)^2}\ ,
\ee
where ${\cal P}(\Phi,V,a,b)$ is a polynomial which is always 
different from zero for any real values of $a$ and $b$.  
In the limiting 
case $a=b=0$, the fundamental equation reduces to that of an ideal gas and 
the curvature vanishes. But the curvature vanishes also in the limiting case 
$a=0$ and $b\neq 0$. This is in agreement with the well-known property \cite{san05c} 
that the constant $a$ is responsible for the non-ideal thermodynamic interactions,
whereas the constant $b$ plays a qualitative role in the description of 
thermodynamic interactions.
Since the main characteristic of an ideal gas is the
absence of thermodynamic interaction between the particles of the gas, we conclude
that the curvature can be used as a measure of this interaction. Any generalization 
of an ideal gas is therefore characterized by a non-zero curvature. 

The scalar curvature for the van der Waals gas is singular for all values of pressure
and volume which satisfy the condition $PV^3-aV+2ab=0$. One can show that this is 
a zero-volume singularity in the sense that it is characterized by the vanishing of
the determinant of the metric $g$. It is a true curvature singularity since it cannot
be removed by a change of coordinates, and we interpret it as an indication that 
our geometric approach is not valid as the singularity is approached. Remarkably, this
limit coincides with the limit of applicability of classical thermodynamics \cite{callen}.
Indeed, the condition $PV^3-aV+2ab=0$ 
for the van der Waals gas is associated with the limit of 
thermodynamic stability, i.e., situations where thermodynamic processes cannot be 
described in the context of equilibrium thermodynamics. This shows that in order 
to avoid the curvature singularity it is necessary to introduce a different approach 
which take into account processes of non-equilibrium thermodynamics. This is, of 
course, a task that requires the application of more general geometric structures, 
and is beyond the scope of the present work.

On the other hand, the limit of thermodynamic stability is associated with 
the appearance of critical points and phase transitions. This opens the 
possibility of analyzing critical behavior and phase transitions of thermodynamic
systems by studying the geometric properties of the curvature singularities. 
In particular, one could try to classify the behavior of the curvature near 
the singularities in terms of divergent polynomials or exponentials, and relate 
them with different types of critical points and phase transitions.

\section{Black holes}
\label{sec:bh}

According to the no-hair theorems of general relativity, 
electro-vacuum black holes are completely described by three
parameters only: mass $M$, angular momentum $J$, and electric charge 
$Q$. These parameters satisfy the first law of black hole 
thermodynamics \cite{bch73}
\be 
dM = T dS + \Omega_H  d J + \phi d Q \ ,
\ee
where $S$ is the entropy, which is proportional to the horizon area, 
$T$ is the Hawking temperature, which is proportional to the surface 
gravity on the horizon, $\Omega_H $ is the angular velocity on the horizon, 
and $\phi$ is the electric potential. In black hole thermodynamics the mass
is considered as the thermodynamic potential so that any function of the
form $M=M(S,J,Q)$ corresponds to a fundamental equation. The most general 
fundamental relation was derived originally by Smarr \cite{smarr73} in the form 
$M^2 = A/16\pi + Q^2/2 + (4\pi/A)( J^2+Q^4/4 )$, where $A$ is the horizon area.
The identification of the entropy as the horizon area in the form $S=kA/4$,
where $k$ is Boltzmann's constant, generates the fundamental equation (we use
units in which $k=1$)
\be
M = \left[ \frac{\pi J^2}{S} + \frac{S}{4\pi}\left(1 + \frac{\pi Q^2}{S}\right)^2\right]^{1/2}\ ,
\label{feqbh}
\ee
which corresponds to the Kerr-Newman black hole \cite{solutions}
\bea
ds^2 = &-&\frac{\Delta - a^2\sin^2\theta}{\Sigma} dt^2 
-\frac{2a\sin^2\theta (r^2+a^2 -\Delta)}{\Sigma} dtd \varphi \nonumber\\
& +& \frac{(r^2+a^2)^2 - a^2\sin^2\theta\, \Delta }{\Sigma} \sin^2\theta d\varphi^2
+\frac{\Sigma}{\Delta} dr^2 + \Sigma d\theta^2 \ , 
\eea
\be 
\Sigma = r^2 + a^2\cos^2\theta\ , \quad \Delta = (r-r_+)(r-r_-)\ ,
 \quad r_\pm = M \pm \sqrt{M^2-a^2-Q^2} \ ,
\ee
where $ a = J/M$ is the specific angular momentum. 

The space of equilibrium states for the Kerr-Newman black hole is $3-$dimensional with
coordinates $S$, $J$, and $Q$. The corresponding metric structure can be derived 
from the fundamental equation (\ref{feqbh}). Then 
\be
g_{_{SS}} = \frac{1}{64\pi^2 S^4 M^2} \left\{-S^4 + \pi^2(4J^2+Q^4)[6S^2
          +8\pi S Q^2 + 3\pi^2(4J^2+Q^4)]\right\} \ , 
\ee
\be
g_{_{SQ}} = \frac{Q}{4S^2M^2}[2\pi J^2 -(\pi Q^2+S)M^2]  \ ,
\quad 
g_{_{SJ}} = -\frac{J}{4S^2M^2}[2\pi M^2 + \pi Q^2+S ]  \ ,
\ee
\be
g_{_{QQ}} = \frac{1}{8\pi S^2 M^2}[4\pi^2 J^2 (3\pi Q^2 + S) + (\pi Q^2+S)^3] \ ,
\ee
\be
g_{_{QJ}} = -\frac{\pi J Q}{2 S^2 M^2}(\pi Q^2+S) \  ,
\quad 
g_{_{JJ}} = \frac{1}{4 S^2 M^2}(\pi Q^2 + S)^2 \ .
\ee
The geometry behind this metric is very rich, and deserves a separate investigation.
In this work we limit ourselves to the analysis of two special cases of this 
general metric, namely, the case $J=0$ which corresponds to the Reissner-Nordstr\"om
metric, and $Q=0$ which corresponds to the Kerr black hole. In the first case, the metric
can be written as
\be
g_{_{RN}}= \frac{1}{2S^2}\left( \pi Q^2 + S\right)\left[\frac{1}{8\pi S}   
\left(3\pi Q^2 - S \right) dS^2 -Q dS dQ + S dQ^2\right]
\ .
\label{metrn}
\ee
In the last section  we noticed that the limit of thermodynamic stability corresponds
to the singularities of the corresponding metric in the space of equilibrium states.
To see if this is also valid in this case we calculate the curvature scalar of the
metric $g_{_{RN}}$, and obtain
\be
R_{_{RN}} = - \frac{8 \pi^2 Q^2 S^2 (\pi Q^2 -3S)}{(\pi Q^2 +S)^3 (\pi Q^2 -S)^2 } \ .
\ee
This scalar presents two critical points. The first one is situated at
$S=\pi Q^2$, which according to the fundamental equation for this case, 
corresponds to an extremal black hole with $M=Q$. The second point 
is at $S =\pi Q^2/3$, where the scalar curvature vanishes identically, 
leading to a flat geometry. This point corresponds to the value 
$M= 2Q/\sqrt{3}$, where the system undergoes a phase transition \cite{davies}.
The behavior of the curvature near these two critical points resembles the 
critical thermodynamic behavior of the Reissner-Nordstr\"om black hole \cite{quev071}.
We conclude that in this particular case GTD correctly 
describes the thermodynamic behavior in terms of geometric concepts.

For the second limiting case with $Q=0$ we obtain the thermodynamic metric
\be
g_{_{Kerr}} = \frac{\pi S}{ S^2 + 4\pi^2 J^2}\left[
\left( \frac{3\pi^2 J^4}{S^4} + \frac{3J^2}{2 S^2} - \frac{1}{16 \pi^2}\right) d S^2
- \frac{ J}{S^3}\left(3 S^2 + 4 \pi^2 J^2 \right) dJ d S + dJ^2\right] \ .
\label{gmkerr}
\ee 
A straightforward calculation shows that the curvature of this metric is zero. 
This would imply that statistical interaction is not present in Kerr black holes,
a result that contradicts the results obtained from standard black hole
thermodynamics, where it was shown that rotating black holes possess a non-trivial 
structure of phase transitions \cite{davies}. 
 
\section{Conclusions}
\label{sec:con}

In this work, we reviewed the main concepts of geometrothermodynamics, and
applied them to completely different thermodynamic systems, namely, 
the van der Waals gas and vacuum black holes. In the first case
we saw that it is possible to reproduce the main aspects of the thermodynamic
behavior by using the curvature of the corresponding metric in the space
of equilibrium states. In the case of black holes we found that only the 
thermodynamics of Reissner-Nordstr\"om black hole  can be reproduced by using
the geometry of the space of equilibrium states. In the case of the Kerr black hole,
however, the geometry is flat and leads to a non-interacting thermodynamic systems,
whereas standard thermodynamics predicts a non-trivial behavior with phase transitions.
Although this last result is negative, it must be emphasized that it was obtained
by using a very particular Legendre invariant metric, which is probably the simplest
consistent generalization of other known metrics that are not invariant with
respect to Legendre transformations. 

We expect that the application of
a different metric will overcome this particular difficulty. In any case, a more
detailed analysis is necessary. In particular, it will be interesting to find a criterion
to select Legendre invariant metrics from the vast number of possible metrics. We expect
that a variational principle applied on all metrics of the space of equilibrium states
will be a plausible criterion. This task is currently under investigation and will be 
presented elsewhere.

\begin{theacknowledgments}
 This work was supported in part by Conacyt, M\'exico, grant 48601, and 
 graduate fellowship 165357.
\end{theacknowledgments}



\begin{thebibliography}{99}
\bibitem{gibbs} J. Gibbs, {\it The collected works}, Vol. 1, Thermodynamics, Yale 
University Press, 1948.

\bibitem{car} C. Charatheodory, {\it Untersuchungen \"uber die Grundlagen der 
Thermodynamik}, Gesammelte Mathematische Werke, Band 2, Munich, 1995.  

\bibitem{her} R. Hermann, {\it Geometry, physics and systems}, Marcel 
Dekker, New York, 1973. 

\bibitem{mru1}R. Mrugala, {\it Geometrical  formulation  of equilibrium   
phenomeno\-logi\-cal ther\-mo\-dynamics}, Rep. Math. Phys. {\bf 14},  419 (1978).

\bibitem{mru2} R. Mrugala, {\it Submanifolds in the thermodynamic phase 
space}, Rep. Math. Phys. {\bf 21}, 197 (1985). 

\bibitem{wei1} F. Weinhold, {\it Metric Geometry of equilibrium thermodynamics I, II,
III, IV, V}, J. Chem. Phys. {\bf 63}, 2479, 2484, 2488, 2496 (1975); {\bf 65}, 558  (1976).

\bibitem{fel1} T. Feldman, B. Andersen, A. Qi and P. Salamon, {\it 
Thermodynamic lengths and intrinsic time scales in molecular relaxation},
Chem. Phys. {\bf 83},  5849 (1985).

\bibitem{gil1} R. Gilmore, {\it Thermodynamic partial derivatives}, J. Chem. Phys.
{\bf 75},  5964 (1981).

\bibitem{sal80} P. Salamon, B. Andersen, P.D. Gait and R.S. Berry, {\it 
The significance of Weinhold's length}, J. Chem. Phys. {\bf 73},   1001 (1980).

\bibitem{sal84} P. Salamon, J. Nulton and E. Ihrig, {\it On the relation 
between entropy and energy relations of electrodynamic length}, J. Chem. Phys.
{\bf 80},  436 (1984).

\bibitem{sal85} P. Salamon, J. Nulton and J. D.  Berry, {\it Length in 
statistical thermodynamics}, J. Chem. Phys. {\bf 82}, 2433 (1985). 


\bibitem{nul85} J. Nulton and P. Salamon, {\it Geometry of the  ideal gas},
Phys. Rev. A {\bf 31},  2520 (1985).

\bibitem{san04} M. Santoro, {\it Thermodynamic length in a two-dimensional 
thermodynamic state space}, J. Chem. Phys. {\bf 121},  2932 (2004).

\bibitem{san05a} M. Santoro, {\it Weinhold  length in an isentropic ideal and 
quasi-ideal gas}, Chem. Phys. {\bf 310},  269 (2005).

\bibitem{san05b} M. Santoro, {\it Weinhold's length in an isochoric thermodynamical system
with constant heat capacity}, Chem. Phys. {\bf 313},  331 (2005).


\bibitem{san05c} M. Santoro and Serge Preston, {\it Curvature of the Weinhold metric
for thermodynamical systems with 2 degrees of freedom}, (2005) arXiv:math-ph/0505010.

\bibitem{rup79} G. Ruppeiner, {\it Thermodynamics: A Riemannian geometric model},
Phys. Rev. A {\bf 20},  1608 (1979).

\bibitem{tormon93} G. F. Torres del Castillo and M. Montesinos-Velasquez,
{\it Riemannian structure of the thermodynamic phase space}, Rev. Mex. F\'\i s.
{\bf 39},  194 (1993).

\bibitem{herlac98} G. Hern\'andez and E. A. Lacomba, {\it Contact Riemannian 
geometry and thermodynamics}, Diff. Geom. and Appl. {\bf 8},  205 (1998).

\bibitem{rup95} G. Ruppeiner, {\it  Riemannian geometry in thermodynamic fluctuation theory},
Rev. Mod. Phys. {\bf 67},  605 (1995); {\bf 68},  313 (1996).

\bibitem{jjk03} D.A. Johnston, W. Janke, and R. Kenna, {\it Information geometry,
one, two, three (and four)}, Acta Phys. Polon. B {\bf 34},  4923 (2003). 

\bibitem{jan04} W. Janke, D.A. Johnston, and R. Kenna, {\it  
Information geometry and phase transitions}, Physica A {\bf 336},  181 (2004).


\bibitem{aman03} J. Aman, I. Bengtsson, and  N. Pidokrajt, {\it 
Geometry of black hole thermodynamics}, Gen. Rel. Grav. {\bf 35}, 1733 (2003). 

\bibitem{aman06a} J. Aman, I. Bengtsson, and  N. Pidokrajt, {\it
Flat information geometries in black hole thermodynamics}, 
Gen. Rel. Grav. {\bf 38}, 1305 (2006). 

\bibitem{aman06b} J. Aman N. Pidokrajt, {\it
 Geometry of higher-dimensional black hole thermodynamics}, 
Phys. Rev. D {\bf 73}, 024017 (2006). 

\bibitem{shen05} J. Shen, R. Cai, B. Wang, and R. Su, {\it 
Thermodynamic geometry and critical behavior of black holes}, (2005) arXiv: gr-qc/0512035.

\bibitem{brohug1} D. C. Brody and L. P. Hughston, {\it Geometry of thermodynamic states},
Phys. Lett. A {\bf 245}, 73 (1998). 

\bibitem{brohug2} D. C. Brody and L. P. Hughston, {\it Geometrisation of statistical 
mechanics}, arXiv:gr-qc/9708032.

\bibitem{brohug3} D. C. Brody and L. P. Hughston, {\it Statistical geometry in quantum mechanics}, arXiv:gr-qc/9701051. 


\bibitem{quezar03} H. Quevedo and R. D. Z\'arate, {\it Differential geometry
and thermodynamics}, Rev. Mex. F\'\i s. {\bf 49 S2},  125 (2003).

\bibitem{quev07} H. Quevedo, {\it Geometrothermodynamics}, J. Math. Phys. 
{\bf 48}, 013506 (2007). 

\bibitem{arnold} V. I. Arnold, {\it Mathematical methods of classical mechanics},
Springer Verlag, New York, 1980.

\bibitem{callen} H. B. Callen, {\it Thermodynamics and an introduction to 
thermostatics}, John Wiley \& Sons, Inc., New York, 1985.



\bibitem{burke} W. L. Burke, {\it Applied differential geometry}, Cambridge University
Press, Cambridge, UK, 1987.


\bibitem{bch73} J. M. Bardeen, B. Carter, and S. W. Hawking, {\it The four laws of 
black hole mechanics}, Commun. Math. Phys. {\bf 31}, 161 (1973).

\bibitem{smarr73} L. Smarr, {\it Mass formula for Kerr black holes}, Phys. Rev. Lett. {\bf 70}, 71 (1973).


\bibitem{solutions} H. Stephani, D. Kramer, M. MacCallum, C. Hoenselaers, and E. Herlt, 
{\it Exact solutions of Einstein's field equations}, Cambridge University Press, 
Cambridge, UK, 2003. 

\bibitem{davies} P. C. W. Davies, {\it Thermodynamics of black holes}, Rep. Prog. Phys.
{\bf 41}, 1313 (1978).
 

\bibitem{quev071} H. Quevedo, {\it Geometrothermodynamics of black holes}, Gen. Rel. Grav.
(2007), in press. 



\end{thebibliography}
\end{document}